\documentclass{article}
\usepackage{latexsym}

\everydisplay{\rm }

\begin{document}
\begin{center}
\LARGE On the Classical Limit  \\ [.125in] of
Spin Network Gravity: Two Conjectures
\\ [.25in] \large Donald E. Neville \footnote{\large Electronic
address: dneville@temple.edu}
\\Department of Physics
\\Temple University
\\Philadelphia 19122, Pa. \\ [.25in]
July 5, 2008 \\ [.25in]
\end{center}

\newcommand{\E}[2]{\mbox{$\tilde{{\rm E}} ^{#1}_{#2}$}}
\newcommand{\A}[2]{\mbox{${\rm A}^{#1}_{#2}$}}
\newcommand{\Np}{\mbox{${\rm N}'$}}
\newcommand{\Etwo}{\mbox{$^{(2)}\!\tilde{\rm E} $}\ }
\newcommand{\Etld }{\mbox{$\tilde{\rm E}  $}\ }
\newcommand{\Vtwosq}{\mbox{$(^{(2)}\!{\rm V})^2 $}}
\newcommand{\Vtwo}{\mbox{$^{(2)}\!{\rm V} $}}
\newcommand{\Vthree}{\mbox{$^{(3)}\!{\rm V} $}}

\newcommand{\bea}{\begin{eqnarray}}
\newcommand{\eea}{\end{eqnarray}}
\newcommand{\be}{\begin{equation}}
\newcommand{\ee}{\end{equation}}
\newcommand{\nn}{\nonumber \\}
\newcommand{\rta}{\mbox{$\rightarrow$}}
\newcommand{\rla}{\mbox{$\leftrightarrow$}}
\newcommand{\eq}[1]{eq.~(\ref{#1})}
\newcommand{\Eq}[1]{Eq.~(\ref{#1})}
\newcommand{\eqs}[2]{eqs.~(\ref{#1}) and (\ref{#2})}
\newcommand{\Eqs}[2]{Eqs.~(\ref{#1}) and (\ref{#2})}
\newcommand{\bra}[1]{\langle #1 \mid}
\newcommand{\ket}[1]{\mid #1 \rangle}
\newcommand{\braket}[2]{\langle #1 \mid #2 \rangle}
\large
\begin{center}
{\bf Abstract}
\end{center}

    Estimates are given of the time scales which
govern spreading of a coherent state wave packet.  The estimates,
based on dimensional
analysis, suggest that spreading should be small
for coherent states with average angular momentum of order 100 or larger.
It is conjectured that
in the classical limit, terms in the Hamiltonian which add a new
vertex will be suppressed, compared to terms which modify the
existing spin network without changing the number of vertices.
\\[.125in]
PACS categories: 04.60, 04.30

\clearpage

\section{Introduction}\label{SecIntroduction}

    In two previous papers, which I will refer to as
papers 1 and 2, I
constructed a set of coherent states for planar gravity waves
\cite{1,2}.  The earlier papers do not propose any Hamiltonian.

    However, there are two (at least) questions about
coherent states
which cannot be answered without knowing the Hamiltonian,
or at least knowing a little bit about the Hamiltonian.
The two questions focus on the rate of spreading
of the coherent state wave packet
(section \ref{SecSpread}), and the possible addition
of non-classical, low spin vertices in the classical
limit (section \ref{SecNewVert}).
The little bit of knowledge comes from dimensional analysis.
The answers given are necessarily tentative, and these
questions should be asked again when more
is known about the Hamiltonian.

    At several points I use the following results,
derived in  appendices
C through E of paper 2. The angles and angular momentum for
the coherent state are
Gaussian distributed, and the
standard deviations of angles are order $1/\sqrt{<L>}$, while
the standard deviations of angular momenta
are order $\sqrt{<L>}$.   $<L>$ is the average,
or peak value of the magnitude of angular momentum for
the coherent state.

    (Coherent states
also contain a parameter t, and appendix
C of paper 2 quotes a value of $1/\sqrt{t}$ for the standard deviation
of L.  However, appendix D shows t
must be order $1/<L>$ in order to minimize the size of
small correction terms.  Therefore the standard deviation
of $1/\sqrt{t}$ is not an exception.  All standard
deviations are order $\sqrt{<L>}$ or $1/\sqrt{<L>}$.

\section{Wave Packet Spreading}\label{SecSpread}

    In paper 2 I
computed explicitly the  matrix elements.
\[
    \bra{u,\vec{p}}(\Etld \mbox{or h})
\ket{u,\vec{p}},
\]
where \Etld and h are densitized
inverse triad and holonomy, and the coherent states are
labeled by a unitary matrix u and a vector $\vec{p}$.
However, there are many other matrix
elements of the form
\[
    \bra{u',\vec{p}'}\Etld\ket{u,\vec{p}}
\]
which
I did not consider. For $u'\approx u$ and  $\vec{p}'\approx
\vec{p}$, the overlap could be large.  Do these matrix elements
play a role?  This is an apparent problem with dealing with an
overcomplete set.

    This problem is perhaps more apparent than real.  In the coherent
state approach to the classical limit, one works only with
diagonal matrix elements, the expectation values
\[
    \bra{u,\vec{p}}(\Etld \, \mbox{or}\, h)\ket{u,\vec{p}}.
\]
One does not
worry about matrix elements to other coherent states, but only
about how rapidly the wave packet $\ket{u,\vec{p}}$ will spread.
Provided the time scale governing the spreading is large, the
expectation values can be used to make predictions.

\subsection{Spreading in the weak field limit}

    In the weak field limit the gravitational Hamiltonian decouples
into a sum of oscillators.  These are especially easy to treat using
coherent state methods.  We can expect something like
\be
    <\delta \Etld(z,T=0)> = A\cos \,(k z +\gamma)
    \label{defgamma}
\ee
for time T = 0; and oscillator packets are known to follow the classical path
exactly for T$>$0 \cite{Glauber}:
\[
    <\delta \Etld(z,T)> = A\cos \,(k z -\omega T +\gamma).
\]
$\delta \Etld$ is the fluctuation of \Etld away from flat space.

    Although the \emph{average} value of oscillator displacement follows
the classical path, this is not enough to guarantee classical behavior.
The fluctuations around the average must also be small.  For the
usual oscillator these fluctuations are determined by $\delta \gamma$,
the uncertainty in the phase $ \gamma$ introduced at \eq{defgamma}.
This uncertainty is connected to the uncertainty in the number of quanta by
$\delta N \delta\gamma \sim 1$, which gives
 $\delta\gamma \sim 1/\sqrt{N}$, since N has
standard deviation $\sqrt{N}$.
In the LQG case presumably $\gamma$ will be a function of the
dimensionless angles $(\alpha,\beta)$ used to define the
unitary matrix u, as well as
the angles defining the unit vector $\hat{p}$. ($\gamma$
could also be a function of a dimensionless ratio of angular momenta,
(average Z component) over $<L>$; but this ratio depends on
the angles already listed.)
The standard deviations for fluctuations in the angles are order
$1/\sqrt{<L>}$.  $<L>$ therefore replaces N; and for L of order
100 or so, the  fluctuations around the classical path should be
small.

\subsection{Spreading in the strong field limit}

    Once the self-interaction of the gravitational packet is
included, it becomes much harder to estimate the rate of spreading.
Consider two familiar quantum mechanical examples, the free
particle and the simple harmonic oscillator.   The spreading of the free
particle wave packet is governed  by the time scale

\be
    T_x = m (\sigma_x)^2/\hbar,
        \label{FreeSpread}
\ee
where m is the mass and $\sigma_x$ is the standard
deviation of the T = 0 Gaussian packet in configuration space.
At the other extreme, the Gaussian packet for the simple harmonic oscillator
does not spread at all \cite{Glauber}.  Evidently the rate of
spreading is highly sensitive
to the details of the energy spectrum \cite{Klauder}.

    The planar case, as well as the general SU(2) case, can be
given an asymptotic region, so can have a
Hamiltonian \cite{Regge, Neville}; and it makes sense to talk about energies E.
If the
energy eigenvalues are evenly spaced, resembling those of the
oscillator, then the likelihood of spreading should be small.

    The Hamiltonian is a surface term.  I will not try to construct this
surface term in spin network theory, but rather will estimate it
using dimensional analysis. The surface term can be rewritten as a
density.  For example for the planar case,

\[
   E= J|^{+\infty}_{-\infty} = \int \partial_z J \,dz,
\]
and the $\partial_z J$ can in turn be expressed in terms of other
fields using the classical equations of motion. One ends up with a
function of the \Etld and h.  I assume this function resembles the
typical terms in the Euclidean Hamiltonian.

\be
    E \sim \epsilon^{ijk} \,Tr(h_{ij} \,h_k \,[h_k^{-1},V] \,)/\kappa^2 .
    \label{EestI}
\ee
(I could also use the
terms in the Hamiltonian which are proportional to the square of the
extrinsic curvature; the order of magnitude estimates would be the same.)
I need to estimate the dependence of this expression on angular momentum
L.  The volume V contains three \Etld operators, integrated over
area, with area eigenvalues of order $L \kappa$.  Therefore volume
V should be order $(L \kappa)^{3/2}$.  However, the commutator of
the volume with holonomy takes the derivative of V with respect to
L; see the discussion of the commutator in paper 2.
Therefore $h_k \,[h_k^{-1},V]$  is order
$\sqrt{L}(\kappa)^{3/2}$.
The remaining holonomies in the Hamiltonian give a result of order
unity when acting on a state.  Therefore the energy grows as the
square root of L.

\be
    E \sim \sqrt{(L \kappa)}/\kappa.
    \label{EestII}
\ee
This resembles the classical expression for the gravitational
Hamiltonian, integral of (curvature) $d^3x /\kappa \simeq$ (large
length)/$\kappa$, except that the large length has been replaced
by the square root of an area eigenvalue.  Also, the
mass and event horizon area of a black hole are connected by
a relation of similar form,
mass $\propto$ square root of area.

    For minimal spreading, the spacing between energy levels should be
as constant as possible, resembling the spacing between levels of the
usual oscillator \cite{Klauder}.  From \eq{EestII} the spacing is order

\be
        \delta E \sim \hbar c \, \delta L /\sqrt{L \kappa}
        \label{EestIII} \ee

I have restored factors of $\hbar$ and c, and given $\kappa$ the dimensions
of a length.  For $\delta L = 1$, define a frequency $\omega$ by

\be
        \hbar \omega := \delta E(\delta L = 1) \sim \hbar c /\sqrt{L \kappa}.
               \label{defw}
\ee

At first glance this result is not what we want: the quantity $\omega$ is not a
constant, independent of L.  However, $\omega$ does not need to be a
constant everywhere; it must be an approximate constant for the range of L
values contained in a coherent state.  Over this range, the fractional
change in $ \omega$ is of order

\be
        \delta \omega/\omega = - \delta L/(2 L),
        \label{dwoverwI} \ee
where now $\delta L$ is the range of L values in the coherent state.  Those
L values are Gaussian distributed with a standard deviation $\sqrt{1/t}$.
Using this standard deviation to estimate
$\delta L$, I get

\be
    \delta \omega/\omega = - 1/(2\sqrt{t} L).
    \label{dwoverwII}
\ee
As mentioned in the introduction, the parameter t must be of
order $t \sim 1/<L>$ in order to minimize
small correction terms.  Inserting this value into \eq{dwoverwII},
and replacing L by $<L>$, I get
\be
    \delta \omega/\omega \sim - 1/(\sqrt{<L>}).
    \label{dwoverwIII}
\ee
Even in the strong field case, a value $<L> \,\geq 100$ should be enough
to drive $\delta \omega$ to zero and prevent
spreading.

    \Eq{dwoverwII} is yet another reason why we cannot make t arbitrarily small:
the spacing between levels would no longer be uniform over the packet, leading
to unacceptable spreading.  It is perhaps relevant that t plays the
role of a standard deviation in spin network theory, and the
time scale $T_x$ for spreading of the free particle packet is also sensitive
to a standard deviation, $\sigma_x$; see \eq{FreeSpread}.

\section{New Vertices in the Classical Limit}\label{SecNewVert}

   It is not immediately clear that the action of the Hamiltonian,
in the classical limit, involves only grasps at vertices with very
high spin.  The Hamiltonian probably contains terms which add
a new vertex to the spin network, a vertex which therefore involves
the minimum spin, spin 1/2.  Consider the Euclidean term in the
Hamiltonian for concreteness.  (Since the remaining terms follow from commutators
involving the Euclidean term, those terms will inherit the properties
of the Euclidean term.)  The spin network version of this term
contains the operators shown in \eq{EestI}.
Consider, for example, the term $h_{ij} = h_{az}$ in this equation.
This is a line integral of the
holonomy along a contour with sides parallel to directions a (= x or y) and z.
To visualize the contour, fold a rectangular sheet of paper into a
cylinder, until the two
opposite edges almost touch.  The contour is given by the boundary of the
sheet, the two almost touching
edges plus the two circular ends.  Align the two almost touching edges with the
z direction of the spin network; align the circular ends with the transverse
a direction.  (The contours in the transverse directions are always closed loops.)  The
contour is not completely closed.  It must be opened, at one corner of the original
sheet of paper, in order to insert the [V,h] commutator.

    The above description is not enough to define the Hamiltonian.  If the
spin network contains vertices $v_1, v_2,\ldots,v_n$ arranged along the z axis,
one must specify how the holonomic contour is positioned with respect to these vertices.
Evidently one of the circular ends, the one containing the [V,h] commutator, must coincide
with one of the vertices, say $v_1$; otherwise the volume operator in [V.h]
will give zero.  As for the other circular end, the
two simplest possibilities  are: that end coincides with the
nearest neighbor, vertex $v_2$ ; or, that end creates a new vertex  with an
associated spin 1/2 loop, at a point $v_{12}$ between $v_1$ and $v_2$.  The $v_{12}$
possibility prompted our earlier speculation that we might have to deal with
non-classical vertices.

    The Hamiltonian probably has to allow for both possibilities, both the
loop ending at
$v_2$, and the loop ending at $v_{12}$.  If the loop only ends
at points $v_{12}$, short of $v_2$,  then no disturbance can propagate
along the lattice from $v_1$ to
$v_2$ and beyond \cite{SmolLR}.  If the loop ends only at the next, already
existing vertex, $v_2$, then the number
of vertices, n, is a good quantum number.  This seems a bit too simple.
For a given fixed total length, the
Hamiltonian should be able to create more vertices, with less
spin per vertex, if the gravitational self-interaction produces
excitations of shorter wavelength.

    If the Hamiltonian can create a new, spin 1/2 vertex, then we
may have to deal with a highly
non-classical action of the Hamiltonian, even in the classical limit.  To
investigate this possibility, I switch to a path integral
point of view,  and estimate the change in action
corresponding to addition of a spin 1/2 loop at $v_{12}$.

    I have to be careful which path integral I choose.  In the
introduction to paper 1, I discussed spin foams
briefly.  The spin foam approach is covariant, uses a path
integral, and
produces a satisfactory evolution operator; but it is not
clear how to extract a canonical Hamiltonian from the
operator \cite{Arnsdorf, Perez}.  If I wish to do a "thought calculation"
involving a path integral, I must
stay on the canonical side of the canonical/covariant divide.

    As pointed out in the previous section, the planar
case has an asymptotic
region, therefore a genuine Hamiltonian.  I use this
Hamiltonian (rather than the spin foam evolution operator)
to construct a path integral.

    This choice of Hamiltonian leads to an action which
is order $\sqrt{L}$, like the Hamiltonian.  Proof:
the action and Hamiltonian
differ by a term $(\Etld/\kappa) \dot{A} \,d^3 x$.  This term  is
order $\sqrt{L}$:
\begin{eqnarray*}
    (\Etld/\kappa) \dot{A} \,d^3 x &\sim & (\Etld/\kappa) dx^a dx^b h_c[H, h_c^{-1}]\epsilon_{abc}\\
                        &\sim & L \,d (\sqrt{L/\kappa})/dL \\
                        &\sim & L \,(1/\sqrt{L \kappa}).
\end{eqnarray*}
On the second line I have assumed that the commutator of
h with the Hamiltonian
is a derivative with respect to L.  $\Box$

    This action must be exponentiated to give a path integral,
and the integration measure contains delta functions which enforce the
constraints.  I assume the integration is confined to
the constraint surface; therefore I can ignore the ghost fields
needed to exponentiate the constraints.  (This is clearly a thought
calculation, rather than an actual calculation!)

Since the action goes as $\sqrt{L}$, the change in
the action, due to a change $\delta L$  in spin, is
\[
    \delta S/S \sim \delta L/L.
\]
The classical limit is attained by minimizing fluctuations of the action.
The L in the denominator suggests that vertices involving small spins will
not contribute significantly in the classical limit.

    In the classical limit, it is plausible that results should approach
those obtained from field theory.  Presumably this requires
that parameters u, $\vec{p}$ vary slowly from vertex to vertex.
(If one requires only that each individual vertex is coherent,
this may not be enough to obtain a classical limit.)  A requirement
of slow variation would again
rule out vertices with spin near 1/2.

\end{document}